\documentclass[aps,prl,twocolumn,preprintnumbers,amsmath,amssymb,longbibliography,prl]{revtex4}
\usepackage{comment} 
\usepackage{amsmath}
\usepackage{amstext}
\usepackage{epsf}
\usepackage{shortvrb} 
\usepackage{amssymb}
\usepackage{bm}
\usepackage{upgreek}
\usepackage{algorithmic}
\usepackage{fancyheadings} 
\usepackage{color}
\usepackage{colortbl}
\usepackage{tcolorbox} 
\usepackage{hyperref} 
\usepackage{caption}
\usepackage{url}

\begin{document}

\title{Fully Relativistic Derivation of the Thermal Sunyaev-Zel'dovich Effect}
\author{Bal\v sa Terzi{\'c}$^{1}$}\thanks{Email: bterzic@odu.edu} 
\author{Geoffrey A.~Krafft$^{1,2}$}
\author{William Clark$^{1}$}
\author{Alexandre Deur$^{1,2}$}
\author{Emerson Rogers$^{1}$}
\author{Brandon Velasco$^{1}$}
\affiliation{
$^1${Department of Physics, Center for Accelerator Science, Old Dominion University, Norfolk, Virginia 23529, USA} \\
$^2${Thomas Jefferson National Accelerator Facility, Newport News, Virginia 23606, USA} \\
}

\date{\today}

\begin{abstract}
We present the first fully and inherently relativistic derivation of the thermal Sunyaev-Zel'dovich effect. This work uses the formalism historically used to compute radiation spectra emerging from inverse Thomson/Compton sources of x-ray radiation. 
Comparing our results to the traditional approach based on relativistically-corrected classical Kompaneets equation, we find small, but systematic differences. Most notable are the modest ($\leq 10\%$) differences in the crossover frequency where the spectral distortion due to the Sunyaev-Zel'dovich effect vanishes, and the energy increase of the distribution at high electron cloud temperatures.
\end{abstract}

\maketitle

When cosmic microwave background (CMB) radiation scatters off hot intracluster electron gas, it results in a small, yet measurable shift in the CMB photons distribution. This effect is called the Sunyaev-Zel'dovich effect (SZE) \citep{SZ_1970,SZ_1972}. There are two components of the SZE: the thermal SZE (tSZE) due to Thomson scattering (special case of Compton scattering when the electron recoil is negligible)
of CMB photons off hot electrons, and the kinetic SZE  due to the cluster moving with respect to the CMB rest frame. The equations describing tSZE were initially derived from the Kompaneets equation \citep{Kompaneets_1957}, a kinetic equation based on non-relativistic electron distribution. After observing that intracluster gas is extremely hot, with $k_B T_e \lesssim 15$~keV \citep{Arnaud_1994,Markevitch_1994,Markevitch_1996,Holzapfel_1997}, it became clear that the classical derivation of the formulae quantifying the tSZE had to be corrected to capture relativistic effects. Over the years, a number of relativistic extensions of the Kompaneets kinetic equation have been reported \citep{Wright_1979,Fabbri_1981,Rephaeli_1995,Rephaeli_1997,Stebbins_1997,Challinor_1998,Sazonov_1998a,Sazonov_1998b,Nozawa_1998,Itoh_1998,Molnar_1999,Dolgov_2001,Itoh_2004,Chen_2021}. All of these generalizations agree for $k_B T_e \lesssim 15$~keV, appropriate for galaxy clusters.

Importantly, the SZE can be used to estimate the present value of the Hubble parameter $H_0$, and as such weigh in on one of the most important problems currently plaguing cosmology---the Hubble tension. It results from the fact that measurements of $H_0$ performed with low-redshift
quantities, e.g., the Type IA supernova \citep{Riess_2016}, consistently yield values larger than measurements from quantities originating at high-redshift, e.g., fits of CMB radiation \citep{Aghanim_2019}. The 5$\sigma$ discrepancy between the two estimates is almost certainly not due to systematic errors in the measurements \citep{Shah_2021,DiValentino_2021}. Adding the SZE as yet another precision method for estimating $H_0$, as was reported in, e.g., Ref.~\citep{Birkinshaw_1994,Reese_2002,Kozmanyan_2019}, may help resolve the Hubble tension.

In this Letter, we present the first fully and intrinsically relativistic derivation of tSZE, based on computing radiation spectra emerging from relativistic Thomson scattering. This approach draws from the considerable body of work done in the context of Thomson/Compton sources of x-ray radiation (for an overview, see Ref.~\citep{Krafft_2010}). Unlike the original Kompaneets equation, the new approach works equally well for up-Comptonization (or inverse Thomson scattering, when energy transfer is from an electron to a photon) as it does for down-Comptonization (or Thomson scattering, when the energy transfer is from a photon to an electron). Ultimately, the tSZE is a mixture of both of these effects. We compare the results from our new, fully relativistic approach to those from the numerical solution of the relativistic Kompaneets equation reported in Ref.~\citep{Itoh_2004}. 

The tSZE effect on the intensity $I$ of the CMB is traditionally quantified by the generalized Kompaneets equation \cite{Carlstrom_2002}:
\begin{equation} \label{old_SZE_1}
\Delta I_{\rm SZE}(\nu) = I_{\rm SZE}(\nu) - I_{\rm CMB}(\nu) = g(x) I_0 y,
\end{equation}
where $\nu$ is the photon frequency, $I_0 = 2(k_B T_{\rm CMB})^3/(hc)^2$, $T_{\rm CMB}$ the temperature of the CMB today, $h$ the Planck constant, $k_B$ the Boltzmann constant, $c$ the speed of light and the $g(x)$ is the frequency dependence in terms of $x\equiv h\nu/(k_B T_{\rm CMB})$:
\begin{align} \label{old_SZE_2}
g(x) & = \frac{x^4 e^x}{(e^x-1)^2} \left(x\frac{e^x+1}{e^x-1}-4\right) F(T_e, x) \nonumber \\
& \equiv g_0(x) F(T_e, x),
\end{align}
where the function $F(T_e, x)$ is the relativistic
correction which is either given as a 
asymptotic expansion in $x$, e.g.~\citep{Itoh_1998},
\begin{align} \label{old_SZE_2b}
F(T_e,x) = 1+\delta_{\rm SZE} (x,T_e),
\end{align}
or as a numerically evaluated generalized relativistic
Kompaneets equation \citep{Itoh_2004}. Non-relativistic Kompaneets equations corresponds to $F(T_e,x) = 1$. Here
$T_e$ is the temperature of the electron gas, and $y$ is the Compton parameter:
\begin{equation} \label{eq_y}
y \equiv \frac{\sigma_{\rm T}}{m_e c^2} \int_0^l n_{e}(x) k_B T_e(x) dx. 
\end{equation}
where $m_e$ is the mass of the electron, and $\sigma_{\rm T}$ is the Thomson cross-section.
For a constant density and temperature ($n_e=const.\equiv n_{e,0}$, $T_e=const.$) spherical intracluster electron gas cloud of diameter $l$,
the Compton parameter is
\begin{equation} \label{eq_y2}
y \approx \frac{k_B T_e}{m_e c^2} n_{e,0} \sigma_{\rm T} l.
\end{equation}
%

\begin{figure}
    \centering
    \includegraphics[width=0.51\textwidth]{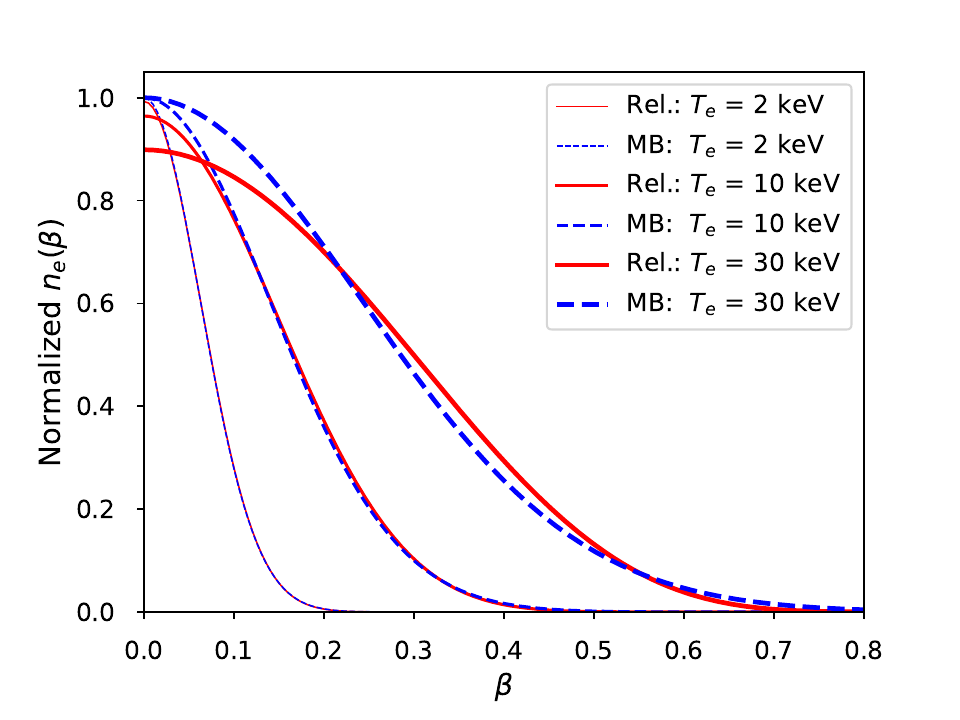}
    \caption{Comparison of the two electron distributions at different temperatures: relativistic (solid lines, Eq.~(\ref{n_e_rel})) and the Maxwell-Boltzmann (dashed lines, Eq.~(\ref{n_e_MB})). Relativistic $\beta=v/c$ is on the $x$-axis. The parameters are $l=2.5$~Mpc, $n_{e,0}=1000~{\rm m}^{-3}$. At each temperature, the two distributions are normalized to $n_e(0)$ of the Maxwell-Boltzmann distribution.}
    \label{fig_dist}
\end{figure}

The CMB photon number density, and the corresponding energy density and intensity (not including anisotropies) are described by the Planck distribution:
\begin{align} \label{eq_nCMB}
n_{\rm CMB}(\nu) & = \frac{8\pi}{c^3} \frac{\nu^2}{e^{h\nu/(k_B T_{\rm CMB})}-1}, \nonumber \\
u_{\rm CMB}(\nu) & \equiv h\nu n_{\rm CMB}(\nu), \nonumber \\
I_{\rm CMB}(\nu) & \equiv \frac{c}{4\pi} h \nu n_{\rm CMB} (\nu).
\end{align}
When integrated over the frequencies $\nu$, Eqs.~(\ref{old_SZE_1}) and (\ref{eq_nCMB}) provide the total intensity difference due to the tSZE and the total intensity of the CMB, respectively:
\begin{align}
\Delta I_{\rm SZE, tot} & = 
\frac{4\pi I_0 y k_B T_{\rm CMB}}{hc}\int g(x) dx, 
\nonumber \\
I_{\rm CMB, tot} & = 
\frac{8 \pi^5 (k_B T_{\rm CMB})^4}{15(hc)^3}.
\end{align}
Therefore, the relative energy shift due to the tSZE is
\begin{equation} \label{old_s}  
s \equiv \frac{\Delta I_{\rm SZE, tot}}{I_{\rm CMB, tot} }.
\end{equation}
At the lowest order, with relativistic effects $\delta_{\rm SZE}$ neglected, $\int_0^\infty g(x) dx = \int_0^\infty g_0(x) dx = 4\pi^4/15$, leading to $s=4y$. Without relativistic effects, this model explicitly conserves the total number of photons because $\int_0^\infty g_0(x)/x dx = 0$. However, with relativistic corrections $\delta_{\rm SZE}$ included as an asymptotic expansion like in Ref.~\citep{Itoh_1998}, each new order correction must separately ensure conservation of photons.

The intracluster medium (ICM) primarily consists of diffused, ionized hydrogen, with trace amount of heavier elements \citep{Mernier_2018}. It has traditionally been modeled as an isothermal sphere of electrons, simplifying the derivation of mass-temperature relationships. However, recent observations and simulations show ICM to be neither isothermal nor perfectly spherical \citep{Lee_2022}. Here we model ICM as hot, relativistic gas of electrons.
The number density of the relativistic electron gas is given by the distribution:
\begin{align} \label{n_e_rel}
n_e (p) d^3 p & \equiv n_{e,0} {\tilde n}_e(v) d^3 p \\
& = n_{e,0} \frac{\hat{\beta}}{4\pi K_2(\hat{\beta})} e^{-\hat{\beta}  \sqrt{1+\gamma^2 \beta^2}} d (\cos{\bar \theta}) d{\bar \phi}  \gamma^5 \beta^2 d \beta, \nonumber
\end{align}
where $p\equiv m_ec \gamma \beta$, $dp=m_ec \gamma^3 d\beta$; $\beta=v/c$ and $\gamma=1/\sqrt{1-\beta^2}$ are the usual relativistic quantities, $({\bar \theta},{\bar \phi})$ are the angles of the electron motion, and $\hat{\beta}=m_ec^2/k_BT_e$. The modified Bessel function $K_2$ is needed to normalize the momentum distribution function. The expression above simplifies to the classical Maxwell-Boltzmann distribution in the limit of small temperatures $T_e$ (small electron velocities $v$):
\begin{align} \label{n_e_MB}
n_e (v) d^3 v & \equiv n_{e,0} {\tilde n}_e(v) d^3 v 
\\
& = n_{e,0} \left(\frac{\hat{\beta}}{2\pi}\right)^\frac{3}{2} e^{-\frac{\hat{\beta} \beta^2}{2}} d (\cos{\bar \theta}) d{\bar \phi} \beta^2 d \beta,
\nonumber
\end{align}
as may be found by taking the asymptotic value for $K_2$. The differences between the relativistic electron distribution in Eq.~(\ref{n_e_rel}) and the Maxwell-Boltzmann distribution in Eq.~(\ref{n_e_MB}) at various temperatures are shown in Fig.~\ref{fig_dist}.

\begin{figure}
    \centering
    \includegraphics[width=0.51\textwidth]{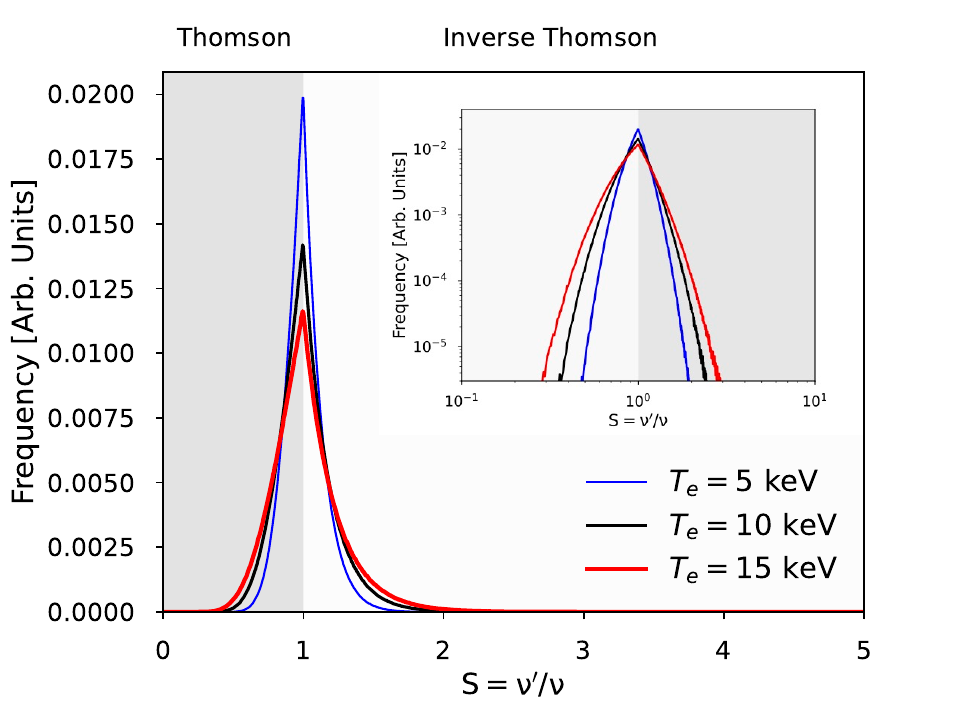}
    \caption{Distribution of values for the scattering function $S$, Eq.~(\ref{eq_scatter_2}). Gray region shows the Thomson scattering of photons, when $\nu' < \nu$ (energy transfers from photons to electrons), while the light region is for the inverse Thomson scattering of photons, when $\nu' > \nu$ (energy transfers from electrons to photons). The parameters are $l=2.5$~Mpc, $n_{e,0}=1000~{\rm m}^{-3}$ and $T_e=5$~keV (thin lines),  $T_e=10$~keV (medium lines), and $T_e=15$~keV (thick lines), with 10 million random samples of the 7-dimensional domain of $S(v, {\bar \theta}, {\bar \phi}, \Theta, \Phi, \theta, \phi)$, binned on 1000 grids. The inset shows the same plot on log-log scale, revealing the power-law tails of the function.}
    \label{fig_Skernel}
\end{figure}

We derive our fully relativistic results by first considering the details of Thomson scattering of a single CMB photon with a single hot electron by enforcing conservation of their relativistic 4-momenta. Let us define the angle that photons make with that line of sight
as $(\Theta, \Phi)$. An electron has a velocity $v$ and angles $({\bar \theta}, {\bar \phi})$.
Then, before the collision, the momenta of the electron and the photon are, respectively:
\begin{align} \label{eq_before}
{\textbf{p}} & = m_e\gamma c(1,\beta \sin{\bar \theta}\cos{\bar \phi}, \beta \sin{\bar \theta} \sin{\bar \phi}, \beta \cos{\bar \theta}), \nonumber \\
{\textbf{k}} & = \frac{h\nu}{c}(1, \sin\Theta \cos \Phi, \sin \Theta \sin \Phi, \cos \Theta).
\end{align}
The collision scatters a photon into angles $(\theta, \phi)$, so:
\begin{align} \label{eq_after}
{\bf{p}}' & = m_e\gamma'(c,p_x',p_y',p_z'),  \\
{\bf{k}}' & \equiv \left(\frac{h \nu'}{c}, {\bm k}'\right) = \frac{h\nu'}{c}(1, \sin \theta \cos \phi, \sin \theta \sin \phi, \cos \theta). \nonumber
\end{align}
The conservation of 4-momentum, 
${\bf{p}} + {\bf{k}}  = {\bf{p}}' + {\bf{k}}'$, relates the
incoming and scattered photon frequencies:
\begin{widetext}
\begin{align} \label{eq_scatter_2}
\nu' (\nu; v, {\bar \theta}, {\bar \phi}, \Theta, \Phi, \theta, \phi) 
& = \nu \frac{
1- \beta (\sin{\bar \theta} \cos{\bar \phi} \sin\Theta \cos\Phi +
\sin{\bar \theta} \sin{\bar \phi} \sin\Theta \sin\Phi +
\cos{\bar \theta} \cos \Theta)
}
{
1-\beta 
(
\sin{\bar \theta} \cos{\bar \phi} \sin\theta \cos\phi +
\sin{\bar \theta} \sin{\bar \phi} \sin\theta \sin\phi +
\cos{\bar \theta} \cos \theta
)
} \nonumber \\
& \equiv \nu S(v, {\bar \theta}, {\bar \phi}, \Theta, \Phi, \theta, \phi). 
\end{align}
\end{widetext}
$S$ is the scattering function which captures the dependence of the scattered photon energy change on the collision kinematics and geometry. There is no nonlinear $h \nu /(\gamma m_e c^2)$ term in the denominator because electron recoil is neglected in Thomson scattering. To make explicit contact with earlier work, in the special case of back-scattering typical for Thomson sources of x-ray radiation, ${\bar \theta}=0$, $\Theta=\pi$, Eq.~(\ref{eq_scatter_2}) reduces to the familiar
\citep{Krafft_2010}
\begin{equation} \label{eq_nuprime}
\nu' = \nu \frac{1+\beta}{1 - \beta \cos \theta}.
\end{equation}
It is important to note that the scattering which occurs as the tSZE includes {\it both} up- and down-Comptonization. Which of the two takes place depends on kinematics and geometry, as quantified by $S$. Figure \ref{fig_Skernel} shows a distribution of values for $S$. Overall, the net effect is  energy increase, as numerically confirmed by $\langle S\rangle > 1$.

The number density of photons scattered by an electron cloud is
%
\begin{align} \label{eq_nuprime}
& n_{\rm s} (\nu')  = \int_0^l \int \int \int n_{\rm CMB}(\nu(\nu')) \frac{d\sigma}{d\Omega} n_e (v) d\Omega_{\bf k} 
d\Omega_{s} d^3v  dx \nonumber \\ 
& = 
l n_{e,0} \int \int \int \int
n_{\rm CMB}(\nu'/S) \frac{d\sigma}{d\Omega} v^2 {\tilde n}_e (v)  d\Omega_{\bf k} 
d\Omega_{s} d\Omega_{\bf p} dv,
\end{align}
%
where $d\Omega_{\bf k}$, $d\Omega_{\bf p}$ and $d\Omega_{s}$ are elements of solid angle for the  photon before scattering, electron before scattering and photon after scattering, respectively, and 
\begin{align} \label{eq_cross_section}
\frac{d{\sigma}}{d\Omega} & = \frac{r_e^2}{\gamma^2 (1-{\bm \beta}\cdot {\hat {\bm k}}')^2}   
\nonumber \\
& \times 
\left[
1-\frac{m_e^2 c^2}{({\bf p}\cdot {\bf k}')^2}
\left(
{\bf k}' \cdot {\bf \epsilon} -
\frac{{\bf p} \cdot {\bf \epsilon}}{{\bf p} \cdot {\bf k}}
{\bf k} \cdot {\bf k}'
\right)^2
\right],
\end{align}
is the Klein-Nishina cross-section \citep{Krafft_2016}, 
${\bm {\hat k}'} = {\bm k}'/|{\bm k}'|$, ${\bf \epsilon}$ is the polarization 4-vector of the incoming photon and $r_e$ is the classical electron radius.
One can show by computing the dot products in Eq.~(\ref{eq_cross_section}) that the dependence on $\nu$ and $\nu'$ stemming from the $k$ and $k'$ terms cancels out---the Klein-Nishina cross-section is independent of the photon energy. This means that the probability $p$ of a photon scattering by the tSZE is 
frequency-independent---the same proportion of CMB photons at all frequencies are scattered by the tSZE. After the tSZE, the total number density of the CMB photons becomes:
\begin{equation} \label{model_n}
n_{\rm SZE} (\nu) = 
(1-p) n_{\rm CMB}(\nu) + n_s(\nu). 
\end{equation}
Computing the number density of the photons
scattered by the tSZE from Eq.~(\ref{eq_nuprime}) requires 7-dimensional numerical integration, which we carry out using Monte-Carlo methods. 

To make contact with the previous work, e.g.~\citep{Itoh_1998,Itoh_2004}, we multiply Eq.~(\ref{model_n}) by
$c h\nu/(4\pi)$ 
\begin{align} \label{model_I}
I_{\rm SZE} (\nu) & = 
(1-p) I_{\rm CMB}(\nu) + I_s(\nu), 
\nonumber \\
I_s(\nu) & = \frac{c}{4\pi} h\nu n_s(\nu),
\end{align}
and 
\begin{align} \label{model_DI}
\Delta I_{\rm SZE} (\nu) & = 
I_{\rm SZE} (\nu)
- I_{\rm CMB}(\nu) \nonumber \\
& =
I_s(\nu) - pI_{\rm CMB}(\nu).
\end{align}

\begin{figure}
    \centering
    \includegraphics[width=0.51\textwidth]{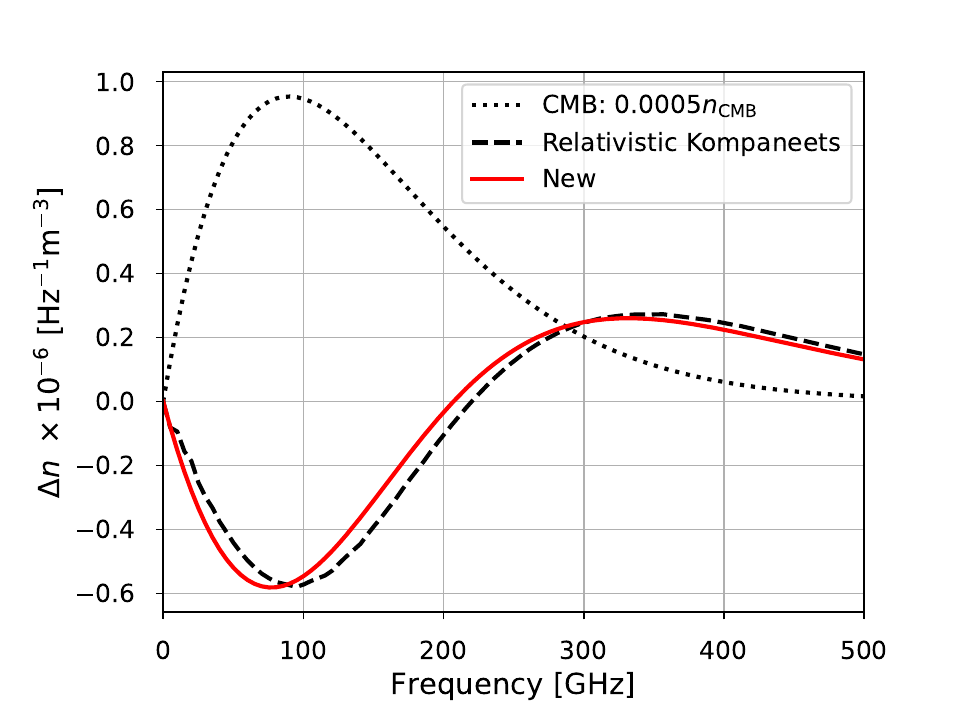}
    \includegraphics[width=0.51\textwidth]{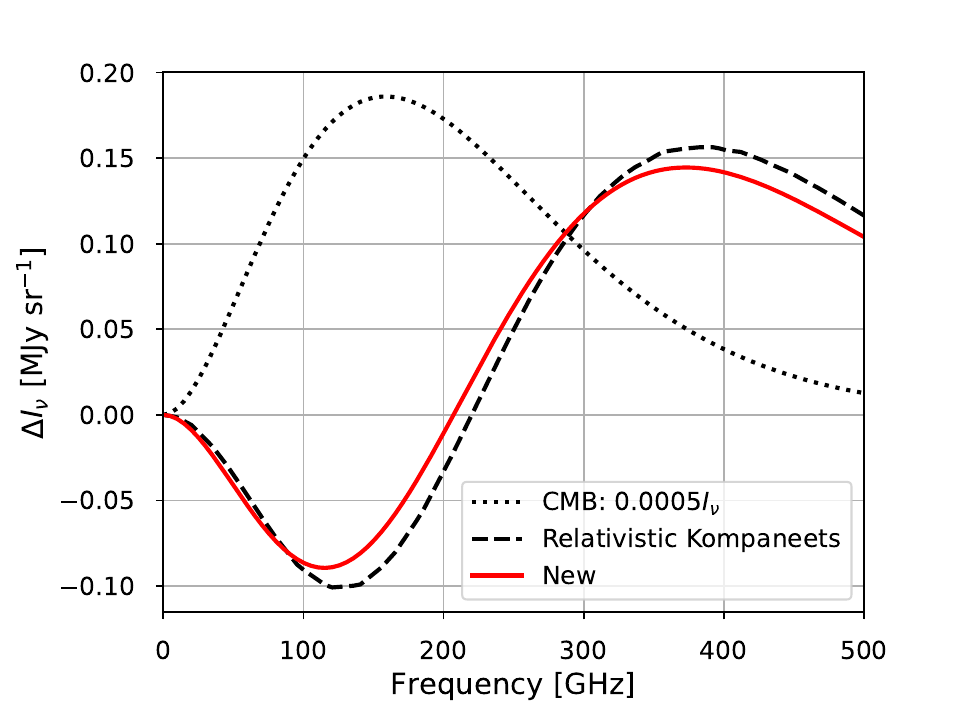}
    \caption{Comparison of the two calculations for the tSZE: the traditional approach with the numerically computed relativistic Kompaneets equation \cite{Itoh_2004} (dashed line), and the fully relativistic approach proposed here (solid line). Number density $n$ is shown on the top and the intensity $I$ on the bottom. The latter is to be compared to Fig.~2 in Ref.~\citep{Carlstrom_2002}. The parameters are $l=2.5$~Mpc, $n_{e,0}=1000~{\rm m}^{-3}$ and $T_e=10$~keV.}
    \label{fig_SZE_PDF}
\end{figure}

The tSZE scatters photons to nearby energies. This redistribution conserves the total number of photons
\begin{align}
\int n_{\rm SZE} (\nu') d\nu' & = 
(1-p) \int n_{\rm CMB}(\nu) d\nu
+ \int n_s(\nu') d\nu' \nonumber \\
& = \int n_{\rm CMB} (\nu) d\nu \equiv n_{\rm CMB,0},
\end{align}
which implies that the probability of a CMB photon scattering by the tSZE is
\begin{align} \label{eq_p}
p \equiv \frac{\int n_s(\nu') d\nu'}{\int n_{\rm CMB} (\nu) d\nu} = \frac{n_{\rm s}}{n_{\rm CMB,0}},
\end{align}
where $n_s \equiv \int n_s(\nu') d\nu'$.
Defining the scattering probability $p$ as in Eq.~(\ref{eq_p}) ensures explicit photon conservation.
Figure \ref{fig_SZE_PDF} shows the tSZE for the numerically computed Kompaneets equation, as reported in Ref.~\citep{Itoh_2004}, and the fully relativistic approach presented here. 

\begin{figure}
    \centering
    \includegraphics[width=0.51\textwidth]{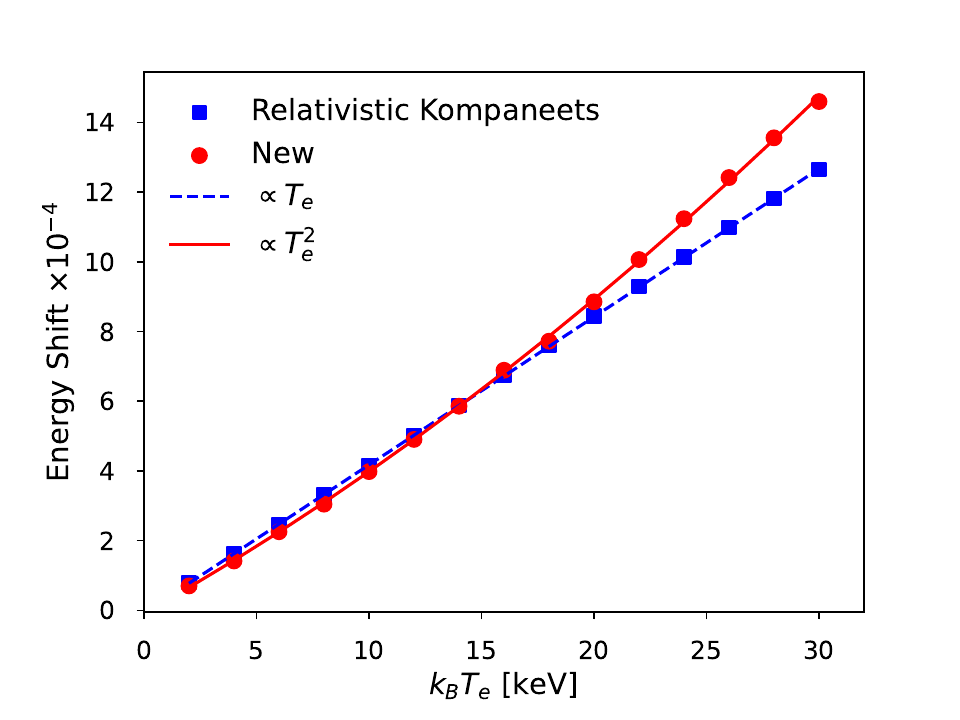}
    \caption{Comparison of the shift predicted in the fully relativistic case (solid circles line, Eq.~(\ref{eq_s})), and the numerically computed relativistic Kompaneets equation (solid squares, Ref.~\citep{Itoh_2004}). The parameters are $l=2.5$~Mpc, $n_{e,0}=1000~{\rm m}^{-3}$, as the electron cloud temperature $T_e$ is varied.}
    \label{fig_SZE}
\end{figure}

The total intensity after accounting for the tSZE comes from the scattered photons and $(1-p)$ original  photons:
\begin{align} \label{SZE_shift}
I_{\rm SZE, tot} & = (1-p) I_{\rm CMB, tot} + I_{s,{\rm tot}},
\end{align}
where 
\begin{align} \label{SZE_shift_2}
I_{s,{\rm tot}} = \frac{c}{4\pi}\int 
h \nu n_{s}(\nu) d\nu.
\end{align}
The corresponding total energy shift for the new approach is then
\begin{align}  \label{eq_s}
\xi  \equiv \frac{I_{\rm SZE, tot} - I_{\rm CMB, tot}}{I_{\rm CMB, tot}}
= \frac{I_{s,{\rm tot}}}{I_{\rm CMB, tot}} - p,
\end{align}
which can be compared to that from the numerically computed Kompaneets equation \citep{Itoh_2004}.
The comparison is shown in Fig.~\ref{fig_SZE}. While  close at lower electron cloud temperatures ($k_B T_e \leq 18$ keV), the two estimates differ noticeably at larger temperatures. 
The range of possible values spanned by the scattering function $S$ is 
$S \in \left[\gamma^{-2} (1+\beta)^{-2},
\gamma^2 (1+\beta)^2 \right]$,   
obtained from Eq.~(\ref{eq_nuprime}). 
From this, we expect the energy shift to scale $\propto (k_B T_e)^2$, consistent with the spectra emerging from inverse Thomson/Compton sources of x-ray radiation. We indeed observe the quadratic dependence on $k_B T_e$ for our $\xi$, while for the traditional estimates it scales linearly with $k_B T_e$. We obtained essentially the same plot when using fifth order asymptotic expansion as reported in Ref.~\citep{Itoh_1998} instead of the numerically computed Kompaneets equation from Ref.~\citep{Itoh_2004}. Our result agrees well with early N-body simulations of the tSZE, that found that for high temperature clusters ($k_B T_e \gtrsim 15~{\rm keV}$) relativistic corrections based on a fifth order expansion of the extended Kompaneets equation seriously underestimate the SZE at high frequencies, with discrepancies in intensity as large as 5~\%, likely leading to $\sim 10$ \% error in estimating the Hubble parameter \citep{Molnar_1999}. 

\begin{figure}
    \centering
    \includegraphics[width=0.48\textwidth]{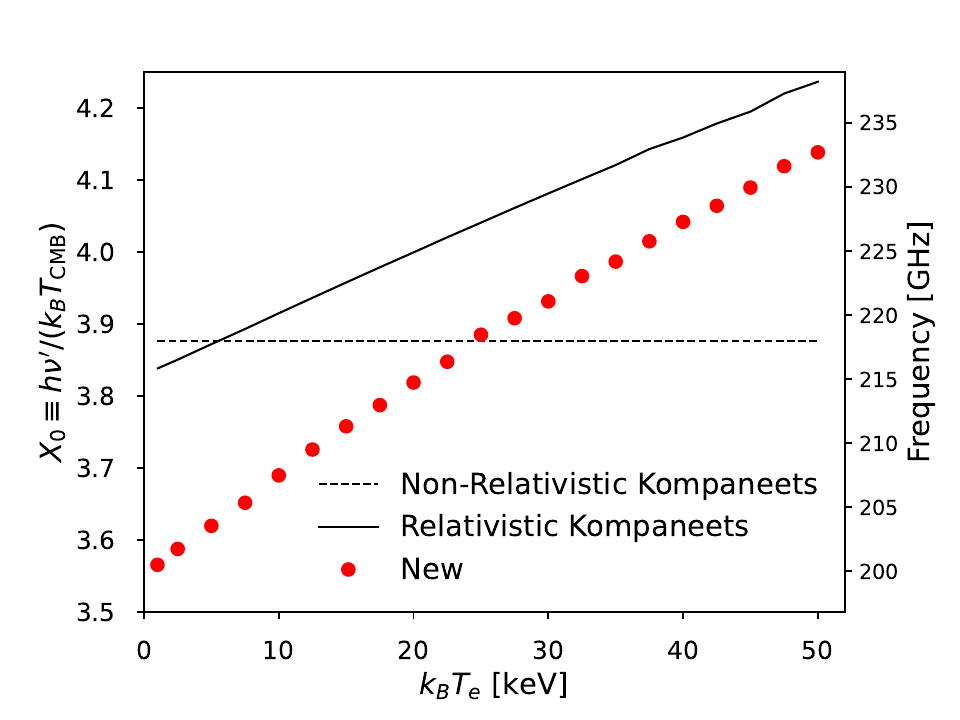}
    \caption{Comparison of the normalized crossover frequency $X_0$ predicted by the new calculation (solid circles), and the numerically computed relativistic Kompaneets equation (solid line, as given in Ref.~\citep{Itoh_2004}) (compare with Fig.~6 from Ref.~\citep{Itoh_1998}). The parameters are $l=2.5$~Mpc, $n_{e,0}=1000~{\rm m}^{-3}$, as the electron cloud temperature $T_e$ is varied. Dashed line denotes the crossover frequency of the non-relativistic Kompaneets equation, 218 GHz, which is independent of the electron gas temperature.}
    \label{fig_crossover}
\end{figure}

Finally, we study the crossover frequency, normalized as $X_0 = h\nu_0/(k_B T_{\rm CMB})$, at which the spectral intensity distortion vanishes. In a non-relativistic model based on traditional Kompaneets equation, the crossover frequency is 218 GHz, independent of electron temperature, optical depth, and all other parameters. Accurate determination of the $X_0$ values is crucial for the study of the SZE \citep{Rephaeli_1995}. In Fig.~\ref{fig_crossover}, we plot the normalized crossover frequency $X_0$ as a function of $k_B T_e$. 

In this Letter, we presented the derivation of the first fully and intrinsically relativistic description of the Thomson scattering process which drives the tSZE. Whereas the present state of the art approach to modeling the tSZE, based on the generalized Kompaneets equation is {\it relativistic by correction}, our new derivation is {\it relativistic by construction}: it is based on conservation of relativistic 4-momenta and relativistic electron distribution. Also by construction, the new calculation ensures photon conservation. At the topmost level, the new approach computes the properties of a single CMB photon relativistically scattered off a single hot electron (a well-understood fundamental process), and then it averages it over the distributions of the two colliding species; Lorentz-transformed Klein-Nishina scattering cross-section controls the likelihood of scattering.
Upon comparing our new result to that based on the generalized Kompaneets equation, we find that the two approaches exhibit non-negligible differences in the shapes of the scattered spectra, the nature of the dependence of the energy shift on the electron cloud temperature and the crossover frequency where the spectral intensity distortion vanishes. Further studies, beyond the scope of the present work, are needed to fully realize the importance of the new results. Detailed comparison of the new approach to the observations can only be done after carefully accounting for the kinetic SZE, systematics, contamination and confusion from astronomical sources.

\vskip5pt
This work is authored by Jefferson Science Associates,
LLC under U.~S.~Department of Energy (DOE) Contract
No.~DE-AC05-06OR23177. The U.~S.~Government retains a
nonexclusive, paid-up, irrevocable, world-wide license to
publish or reproduce this manuscript for U.S. Government
purposes. B.~T.~acknowledges the support of NSF CAREER
Grant No.~1847771.


\begin{thebibliography}{10}

\bibitem{SZ_1970}
R.~A. {Sunyaev} and Ya.~B. {Zeldovich}.
\newblock {\em Comments on Astrophysics and Space Physics}, 2:66, 1970.

\bibitem{SZ_1972}
R.~A. {Sunyaev} and Ya.~B. {Zeldovich}.
\newblock {\em Comments on Astrophysics and Space Physics}, 4:173, 1970.

\bibitem{Kompaneets_1957}
A.~S. Kompaneets.
\newblock {\em Soviet Journal of Experimental and Theoretical Physics},
  4(5):730--737, 1957.

\bibitem{Arnaud_1994}
K.~A. {Arnaud}, R.~F. {Mushotzky}, H.~{Ezawa}, Y.~{Fukazawa}, T.~{Ohashi},
  M.~W. {Bautz}, G.~B. {Crewe}, K.~C. {Gendreau}, K.~{Yamashita}, Y.~{Kamata},
  and F.~{Akimoto}.
\newblock {\em Astrophysical Journal Letters}, 436:L67, 1994.

\bibitem{Markevitch_1994}
M.~{Markevitch}, K.~{Yamashita}, A.~{Furuzawa}, and Y.~{Tawara}.
\newblock {\em Astrophysical Journal}, 436:L71, 1994.

\bibitem{Markevitch_1996}
M.~{Markevitch}, R.~{Mushotzky}, H.~{Inoue}, K.~{Yamashita}, A.~{Furuzawa}, and
  Y.~{Tawara}.
\newblock {\em Astrophysical Journal}, 456:437, 1996.

\bibitem{Holzapfel_1997}
W.~L. {Holzapfel}, M.~{Arnaud}, P.~A.~R. {Ade}, S.~E. {Church}, M.~L.
  {Fischer}, P.~D. {Mauskopf}, Y.~{Rephaeli}, T.~M. {Wilbanks}, and A.~E.
  {Lange}.
\newblock {\em Astrophysical Journal}, 480(2):449--465, 1997.

\bibitem{Wright_1979}
E.~L. {Wright}.
\newblock {\em Astrophysical Journal}, 232:348--351, 1979.

\bibitem{Fabbri_1981}
R.~{Fabbri}.
\newblock {\em Astrophysics and Space Science}, 77(2):529--537, 1981.

\bibitem{Rephaeli_1995}
Y.~{Rephaeli}.
\newblock {\em Astrophysical Journal}, 445:33, 1995.

\bibitem{Rephaeli_1997}
Y.~Rephaeli and D.~Yankovitch.
\newblock {\em Astrophysical Journal}, 481(2):L55, 1997.

\bibitem{Stebbins_1997}
A.~Stebbins.
\newblock {\em arXiv preprint astro-ph/9705178}, 1997.

\bibitem{Challinor_1998}
A.~Challinor and A.~Lasenby.
\newblock {\em Astrophysical Journal}, 499(1):1, 1998.

\bibitem{Sazonov_1998a}
S.~Y. {Sazonov} and R.~A. {Sunyaev}.
\newblock {\em Astrophysical Journal}, 508(1):1--5, 1998.

\bibitem{Sazonov_1998b}
S.~Yu. {Sazonov} and R.~A. {Sunyaev}.
\newblock {\em Astronomy Letters}, 24(5):553--567, 1998.

\bibitem{Nozawa_1998}
S.~Nozawa, N.~Itoh, and Y.~Kohyama.
\newblock {\em Astrophysical Journal}, 507(2):530, 1998.

\bibitem{Itoh_1998}
N.~{Itoh}, Y.~{Kohyama}, and S.~{Nozawa}.
\newblock {\em Astrophysical Journal}, 502(1):7--15, 1998.

\bibitem{Molnar_1999}
S.~M Molnar and M.~Birkinshaw.
\newblock {\em Astrophysical Journal}, 523:78, 1999.

\bibitem{Dolgov_2001}
A.~D. Dolgov, S.~H. Hansen, S.~Pastor, and D.~V. Seikoz.
\newblock {\em Astrophysical Journal}, 554:74, 2001.

\bibitem{Itoh_2004}
N.~Itoh and S.~Nozawa.
\newblock {\em Astronomy \& Astrophysics}, 417:827, 2004.

\bibitem{Chen_2021}
{X.~Chen}, {X.~Zhang}, {H.~Gao}, {C.~Han}, and {D.~Liu}.
\newblock {\em Astronomy \& Astrophysics}, 650:A74, 2021.

\bibitem{Riess_2016}
A.~G. Riess et~al.
\newblock {\em Astrophysical Journal}, 826(1):56, 2016.

\bibitem{Aghanim_2019}
N.~Aghanim et~al.
\newblock {\em Astronomy \& Astrophysics}, 641:A5, 2020.

\bibitem{Shah_2021}
P.~Shah, P.~Lemos, and O.~Lahav.
\newblock {\em Astronomy \& Astrophysics Review}, 29:1, 2021.

\bibitem{DiValentino_2021}
E.~Di~Valentino, O.~Mena, S.~Pan, L.~Visinelli, W.~Yang, A.~Melchiorri, D.~F.
  Mota, A.~G. Riess, and J.~Silk.
\newblock {\em Classical and Quantum Gravity}, 38(15):153001, 2021.

\bibitem{Birkinshaw_1994}
M.~{Birkinshaw} and J.~P. {Hughes}.
\newblock {\em Astrophysical Journal}, 420:33, 1994.

\bibitem{Reese_2002}
E.~D. Reese, J.~E. Carlstrom, M.~Joy, J.~J. Mohr, L.~Grego, and W.~L.
  Holzapfel.
\newblock {\em Astrophysical Journal}, 581(1):53, 2002.

\bibitem{Kozmanyan_2019}
{A.~Kozmanyan}, {H.~Bourdin}, {P.~Mazzotta}, {E.~Rasia}, and {M.~Sereno}.
\newblock {\em Astronomy \& Astrophysics}, 621:A34, 2019.

\bibitem{Krafft_2010}
G.~A. Krafft and G.~Priebe.
\newblock {\em Reviews of Accelerator Science and Technology}, 03(01):147--163,
  2010.

\bibitem{Carlstrom_2002}
J.~E. Carlstrom, G.~P. Holder, and E.~D. Reese.
\newblock {\em Annnual Reviews of Astronomy and Astrophysics}, 40:643--680,
  2002.

\bibitem{Mernier_2018}
F.~{Mernier}, N.~{Werner}, J.~{de Plaa}, J.~S. {Kaastra}, A.~J.~J. {Raassen},
  L.~{Gu}, J.~{Mao}, I.~{Urdampilleta}, and A.~{Simionescu}.
\newblock {\em Monthly Notices of the Royal Astronomical Society},
  480(1):L95--L100, 2018.

\bibitem{Lee_2022}
E.~Lee, D.~Anbajagane, P.~Singh, J.~Chluba, D.~Nagai, S.~T. Kay, W.~Cui,
  K.~Dolag, and G.~Yepes.
\newblock {\em Monthly Notices of the Royal Astronomical Society},
  517(4):5303--5324, 2022.

\bibitem{Krafft_2016}
G.~A. {Krafft}, E.~{Johnson}, K.~{Deitrick}, B.~{Terzi{\'c}}, R.~{Kelmar},
  T.~{Hodges}, W.~{Melnitchouk}, and J.~R. {Delayen}.
\newblock {\em Physical Review Accelerators and Beams}, 19(12), 2016.

\end{thebibliography}
\end{document}